\documentclass[12pt,a4paper]{article}

\newcommand{\cL}{{\cal L}}
\newcommand{\cM}{{\cal M}}
\newcommand{\Ord}[1]{{\cal O}\left(#1\right)}

\usepackage{epsfig}
\epsfxsize=\textwidth
\renewcommand{\includegraphics}[1]{\epsfbox{#1}}

\begin{document}
\title{\sf
Similarity, attraction and initial conditions in an example of nonlinear
diffusion
}
\author{S.A.~Suslov\thanks{Dept of Mathematics \& Computing,
University of Southern Queensland, Toowoomba, Qld 4352, Australia.
E-mail: \texttt{ssuslov@usq.edu.au}}
\and
A.J.~Roberts\thanks{Dept of Mathematics \& Computing,
University of Southern Queensland, Toowoomba, Qld 4352, Australia.
E-mail: \texttt{aroberts@usq.edu.au}}}
\date{February 25, 1998}

\maketitle

\begin{abstract}
Similarity solutions play an important role in many fields of science.
The recent book of Barenblatt \cite{Barenblatt96} discusses many
examples.
Often, outstanding unresolved issues are whether a similarity solution
is dynamically attractive, and if it is, to what particular solution
does the system evolve.
By recasting the dynamic problem in a form to which centre manifold
theory may be applied, based upon a transformation by
Wayne~\cite{Wayne94}, we may resolve these issues in many cases.
For definiteness we illustrate the principles by discussing the
application of centre manifold theory to a particular nonlinear
diffusion problem arising in filtration.
Theory constructs the similarity solution, confirms its relevance, and
determines the correct solution for any compact initial condition.
The techniques and results we discuss are applicable to a wide range
of similarity problems.
\end{abstract}

\tableofcontents

\section{Introduction}
Consider the nonlinear diffusion problem with a step in the diffusivity
discussed by Barenblatt \cite[\S3.2]{Barenblatt96} which in
nondimensional form is
\begin{equation}
        \theta_t=\left\{
        \begin{array}{ll}
                \theta_{xx}\,, & \theta_t\geq 0  \\
                (1+\epsilon)\theta_{xx}\,, & \theta_t\leq 0
        \end{array}\right.\,,
        \label{Ebaren}
\end{equation}
where $\theta(x,t)$ is the evolving concentration of some spatially
distributed substance.
Such a problem, with its nonlinear step in the diffusivity, arises in
theory of filtration of an elastic fluid in an elasto-plastic porous
media (see the discussion in \cite[\S3.2.1]{Barenblatt96}).
It describes the diffusion in one spatial dimension $x$ which is
assumed here to be effectively of infinite extent.

We write and analyse~(\ref{Ebaren}) as a perturbation of the basic
linear diffusion problem, namely
\begin{equation}
        \theta_t=\theta_{xx}+f(\theta,\epsilon)\,,
        \label{Eprob}
\end{equation}
where, since $\theta_t$ has the same sign as $\theta_{xx}$,
\begin{equation}
        f=\left\{
        \begin{array}{ll}
                0\,, & \theta_{xx}\geq 0  \\
                \epsilon\theta_{xx}\,, & \theta_{xx}\leq 0
        \end{array}\right.\,.
        \label{Enon}
\end{equation}
The term $f(\theta,\epsilon)$ acts as a nonlinear perturbation to the
basic diffusion of
\begin{equation}
        \theta_t=\theta_{xx}
        \label{Ebas}
\end{equation}
on an infinite domain.  Of course $\epsilon$ need not be small but we
shall treat it so.

We apply centre manifold theory to help understand and solve this
problem.
But on the infinite spatial domain there is no clear cut centre
eigenspace for~(\ref{Ebas}).
However, following Wayne \cite{Wayne94,Wayne96} we transform the problem to
one of seeking $\phi(\xi,\tau)$ where
\begin{equation}
        \tau=\log {t}\,,\quad
        \xi=\frac{x}{\sqrt{t}}\,,\quad
        \theta=\frac{1}{\sqrt t}\phi(\tau,\xi)\,.
        \label{Etrans}
\end{equation}
Then the dependence upon the scaled space variable $\xi$ causes the Gaussian
spread from a point release,
\begin{equation}
        \theta=\frac{a}{2\sqrt{\pi t}}e^{-x^2/(4t)}\,,
        \label{Egauss}
\end{equation}
to correspond to a fixed point of the dynamics for $\phi$, namely
\begin{equation}
        \phi_*=\frac{a}{2\sqrt{\pi}}e^{-\xi^2/4}\,.
        \label{Efix}
\end{equation}
Also, the algebraic decay in $t$ from any compact release to the
Gaussian~(\ref{Egauss}) transforms to an exponentially quick decay in
$\tau$ to the fixed point~(\ref{Efix}).
Centre manifold theory is applied in Section~2 to justify the
self-similar Gaussian~(\ref{Egauss}) as a valid approximation to the
long-term dynamics of the non-constant diffusivity
problem~(\ref{Ebaren}).
Then the centre manifold analysis, as extended in Section~3,
determines that the amplitude $a$ of the decaying Gaussian evolves
like
\begin{equation}
        a\approx a_0 t^{-\epsilon/\sqrt{2\pi e}}
        \label{Esimsoln}
\end{equation}
in accordance with the result reported by Barenblatt for
$\epsilon\neq0$.
In addition to this confirmation of earlier results, centre manifold
theory \cite{Carr} immediately guarantees the attraction of the
similarity solution.
That is, this approach easily establishes the relevance of the
similarity solution to the long-term dynamics of this nonlinear
diffusion and we expect it to be able to analogously justify the
relevance of similarity solutions for other problems.

The amplitude of the spreading Gaussian not only decays in time, it
also is a function of the initial distribution $\theta(x,1)$ of the
substance (note that the initial release is assumed to occur at $t=1$
corresponding to the transformed time $\tau=0$).
Qualitatively, the long term behaviour is similar for all initially
compact releases.
However, the specific evolution of the model does depend on the
specific initial conditions.
In other words, we need to determine $a_0$ in~(\ref{Esimsoln}).
Naively we may expect that the total amount of substance in the model,
given by $a$ in~(\ref{Egauss}),  will be the same as that at the
instant of release and so use
\begin{equation}
        a_0=\int_{-\infty}^\infty \theta(x,1)\,dx\,.
        \label{Enaiveic}
\end{equation}
However, this is only a leading order approximation and needs
correction depending upon other details of the release distribution
$\theta(x,1)$.
The corrections cannot be determined by scaling law arguments, but
require a knowledge of the dynamics of approach to the similarity
solution.
Recently developed theory \cite{Roberts89b,Roberts97b} is used in
Section~4 to determine the proper choice of the initial conditions for
the model amplitude $a$.

For any given release of substance, the assumed origin of space-time
may not be the best location for the origin of the similarity solution.
In Section~5 we show how the translational degrees of freedom in the
coordinate system can be incorporated into the model for it to
represent better the solution of the original diffusion problem.
Numerical solutions reported in Section~6 confirm the effectiveness of
the correct choice of $a_0$ as well as of time and space origins of
the model.

Finally we comment that the example discussed in detail here is just
one of a wide class of nonlinear advection-reaction-diffusion problems.
Centre manifold theory may be successfully applied to many of these
problems and not only create the similarity solution, but also justify
its relevance as an attractive manifold, and determine the correct
initial amplitude for the similarity solutions.
One class of nonlinear reaction-diffusion problems was similarly
analysed by Gene Wayne \cite{Wayne94}.
Some of the similarity solutions of the nonlinear advection diffusion
problems discussed by Doyle and Englefeld \cite{Doyle90} are also
amenable to this centre manifold approach.

\section{Similarity solutions form a centre manifold}

Now investigate the centre manifold analysis in more detail.  The
transformation~(\ref{Etrans}) changes~(\ref{Eprob}) to
\begin{equation}
        \phi_\tau=\cL\phi+f(\phi,\epsilon)\,,
        \label{Ephit}
\end{equation}
where the linear operator
\begin{equation}
        \cL\phi=\phi_{\xi\xi}+\frac{1}{2}\xi\phi_\xi+\frac{1}{2}\phi\,.
        \label{Elphi}
\end{equation}
Adjoin the trivial equation
\begin{equation}
        \epsilon_\tau=0\,.
        \label{Eepst}
\end{equation}
Then observe that for $\epsilon=0$ the Gaussian~(\ref{Efix}) describes
a fixed point of~(\ref{Ephit})--(\ref{Eepst}) for all amplitudes $a$.
Thus the centre manifold we construct will be global in $a$ and local
only in $\epsilon$.
Now the linear operator $\cL$ has a spectrum of
\begin{equation}
        \lambda=-n/2\,,
        \quad n=0,1,2,\ldots\,.
        \label{Espec}
\end{equation}
This is straightforwardly shown by looking for eigensolutions in the
form $e^{\lambda_n \tau-\xi^2/4}H_n(\xi)$, where $H_n$ are Hermite
polynomials \cite{AbramowitzStegan}.  With two zero eigenvalues, one
from~(\ref{Espec}) and one trivially from~(\ref{Eepst}), and the rest
strictly negative, centre manifold theory asserts there exists a two
dimensional centre manifold for~(\ref{Ephit})--(\ref{Eepst}), $\cM_c$,
which is exponentially attractive to nearby trajectories.

Thus by Theorem 2 in \cite[p.4]{Carr}, centre manifold theory
immediately proves the attraction to the asymptotic similarity
solution, albeit only for small enough $\epsilon$.
(Contrast the ease of obtaining this result with Barenblatt's stability
analysis \cite[\S8.3.2]{Barenblatt96}.)
In agreement with Barenblatt's equation~(8.67), from the
spectrum~(\ref{Espec}), we immediately deduce that the longest-lasting
transient in the approach to the similarity solution will be of
relative magnitude approximately $e^{-\tau/2}=1/\sqrt {t}$.

We now approximate $\cM_c$, parameterized by $a$ and $\epsilon$, and
the evolution thereon by
\begin{equation}
        \phi=a(\tau)\left[\psi_0(\xi)+\epsilon \psi_1(\xi)+\epsilon^2
\psi_2(\xi)+\Ord{\epsilon^3}\right]\,,\quad
        \mbox{where}\quad \psi_0=\frac{e^{-\xi^2/4}}{2\sqrt{\pi}} \,,
        \label{Ecm1}
\end{equation}
\begin{equation}
        \mbox{s.t.} \quad
        \dot a=ag=a\left[\epsilon g_1+\epsilon^2 g_2+\Ord{\epsilon^3}\right]
        \label{Ecm1a}
\end{equation}
($\psi_0$ is normalised such that
$\int_{-\infty}^\infty\psi_0\,d\xi=1$ and the overdot denotes
$d/d\tau$).
Substituting (\ref{Ecm1}) and (\ref{Ecm1a}) into~(\ref{Ephit}) and
equating all terms of $\Ord{\epsilon}$ we need to solve
\begin{equation}
  \cL\psi_1=\psi_{0}g_1-D_{\xi_0}\psi_0\,,
  \label{Efirst}
\end{equation}
where for any $s$
\begin{equation}
        D_s=\left\{
        \begin{array}{ll}
                0\,, & \xi\notin[-s,s]  \\
                \frac{\partial^2}{\partial\xi^2}\,,& \xi\in [-s,s]
        \end{array}\right. \,.
\end{equation}
Here $\xi_0=\sqrt{2}$ is such that
$\psi_{0\xi\xi}(-\xi_0)=\psi_{0\xi\xi}(\xi_0)=0$.
But $\cL$ is singular as it has a zero eigenvalue; so we choose $g_1$
to put the remaining terms in the range of $\cL$---this is the
solvability condition.
In order to do this we take the inner product of equation
(\ref{Efirst}) with the solution $z$ of the adjoint problem
\begin{equation}
        \cL^\dag z\equiv z_{\xi\xi}-\frac{1}{2}\xi z_{\xi}=0\,,
        \label{Eladj}
\end{equation}
where the adjoint is obtained using the obvious inner product
\begin{equation}
        \left\langle u,v \right\rangle\equiv\int_{-\infty}^\infty uv\,d\xi\,.
        \label{Edot}
\end{equation}
For a reason discussed later in the paper we normalise the adjoint
eigenvector such that $\left\langle z,\psi_0 \right\rangle=1$.
It is straightforward to check that the adjoint eigenvector satisfying this
normalisation
is $z=1$.

Finally, applying the solvability condition we find that
\begin{equation}
        g_1=2\psi_{0\xi}(\xi_0)=-\frac{1}{\sqrt{2\pi e}}\,.
        \label{Egd}
\end{equation}
(As usual, we do not need to find $\psi_1$ to determine the leading
order evolution.)  The leading order centre
manifold model $\dot a\approx-\epsilon a/\sqrt{2\pi e}$ then has solution
\begin{equation}
        a=a_0e^{-\epsilon\tau/\sqrt{2\pi e}}
        =a_0t^{-\alpha/2}\,,
        \quad\mbox{where}\quad
        \alpha=\epsilon\sqrt{\frac{2}{\pi e}}
        \label{Epow}
\end{equation}
in agreement with Barenblatt \cite[pp175--6]{Barenblatt96}. 
The constant $a_0$ is determined by the initial conditions for the full
original problem and will be determined in Section~4.

\section[The next order correction]%
{The next-order correction matches earlier results}

Before proceeding to the next order approximation for the evolution on the
centre manifold we need to find $\psi_1$.

\begin {figure} [tbp]
  \centerline{\includegraphics{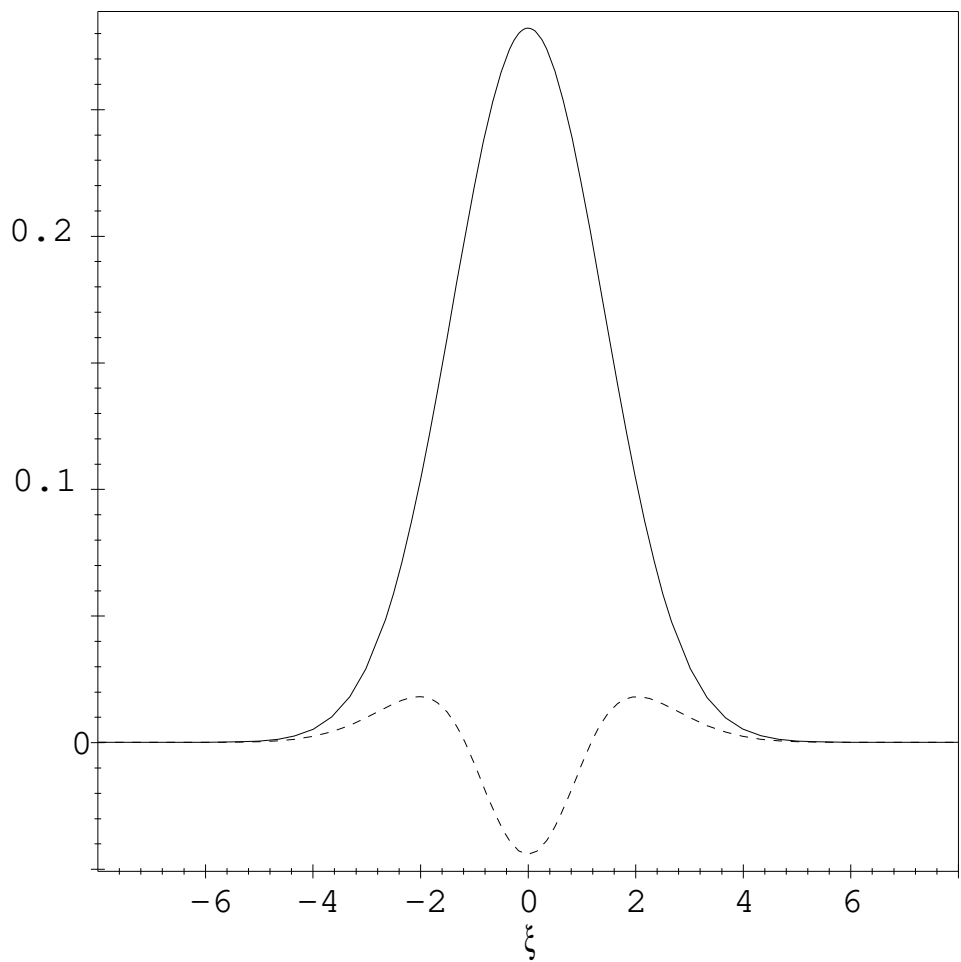}}
  \caption{Solutions $\psi_0(\xi)$ (solid line) showing the Gaussian
  shape of the basic similarity solution, and $\psi_1(\xi)$ (dashed
  line) showing that the Gaussian is flattened and broadened by the
  nonlinear diffusion.}
  \label{f1}
\end{figure}

Since the operator $\cL$ is singular the solution is
not unique and we are free to impose one additional condition on the solution
to fix it. It is convenient to require that
\begin{equation}
        \int_{-\infty}^\infty \psi_1\,d\xi=0\,.
        \label{f1norm}
\end{equation}
Physically this implies that the total amount of the diffused substance is
given completely by the leading order approximation of the solution, and as
$\int_{-\infty}^\infty\psi_0\,d\xi=1$, the total amount is simply $a$.
Under this condition the continuous, up to the second derivative, solution
to (\ref{Efirst}) becomes

\begin{equation}
  \begin{array}{rl}
    \psi_1&=e^{-\xi^2/4}\left\{c_3+
      \frac{i}{2\sqrt{2e}}
      \left(\mbox{erf}(\frac{|\xi|}{2})-1\right)\,
      \mbox{erf}\left(\frac{i|\xi|}{2}\right) \right. \\ &
      \left.-\frac{i}{2\sqrt{2\pi e}}
      \int_0^{\xi} \,\mbox{erf} \left (\frac{iy}{2}\right)
      e^{-y^2/4}dy \right.\\&
      +\left [\frac{\xi^2-2}{8\sqrt{\pi}}+\frac{i}{2\sqrt{2e}}
          \left(\mbox{erf}\left(\frac{i|\xi|}{2}\right)-
            \mbox{erf}\left(\frac{i}{\sqrt 2}\right)\right) \right]\\&
        \left.\quad\times(H(\xi+\xi_0)-H(\xi-\xi_0)) \right\},
\end{array}
        \label{phi1}
\end{equation}
where $H$ denotes the Heaviside function and
\begin{eqnarray}
  c_3=\frac{1}{2\pi\sqrt{2 e}}\left[1+i\pi \,
    \mbox{erf}\left(\frac{1}{\sqrt2}\right)\,
    \mbox{erf}\left(\frac{i}{\sqrt2}\right)
    -i\sqrt\pi\left(I_1+I_2-\frac{I_3}{\sqrt\pi}\right)\right]\\
 \approx-0.1076980691\,. \nonumber
\end{eqnarray}
The integrals entering the definition of $c_3$ are:
\begin{equation}
  I_1=\int_0^{\xi_0} e^{-\frac{\xi^2}{4}}\,\mbox{erf}\left(\frac{\xi}{2}\right)
  \,\mbox{erf}\left(\frac{i\xi}{2}\right)\,d\xi\
\approx0.2262196880i\,,
\end{equation}
\begin{equation}
  I_2=\int_{\xi_0}^\infty e^{-\frac{\xi^2}{4}}\left[\mbox{erf}
  \left(\frac{\xi}{2}\right)-1\right]\,\mbox{erf}\left(\frac{i\xi}{2}\right)
  \,d\xi
  \approx -0.1358229603i\,,
\end{equation}
\begin{equation}
  I_3=\int_0^\infty e^{-\frac{\xi^2}{4}}\int_0^\xi e^{-\frac{y^2}{4}}
\,\mbox{erf}\left(\frac{iy}{2}\right)\,dy\,d\xi\approx 0.6931471806i\,.
\end{equation}
As expected the first order correction, $\psi_1$, is an even function
of $\xi$, see Figure~\ref{f1}.

Let $\psi_{\xi\xi}(\bar{\xi})=0$. Then $\bar{\xi}=\xi_0+\epsilon\xi_1+
\Ord{\epsilon^2}$ where, as is deduced from (\ref{Ecm1}) and (\ref{phi1}),
\begin{equation}
  \xi_1=-\frac{\psi_{1\xi\xi}(\xi_0)}{\psi_{0\xi\xi\xi}(\xi_0)}\approx
  0.5665706981.
  \label{xi1}
\end{equation}
Collecting terms of $\Ord{\epsilon^2}$ we obtain
\begin{equation}
  \cL\psi_2=\psi_{1} g_1+\psi_{0} g_2
  -\left( D_{\xi_0+\epsilon\xi_1}
  -D_{\xi_0} \right) \psi_0
  -D_{\xi_0}\psi_1\,.
  \label{lf2}
\end{equation}
Similarly to the previous section, the application of the solvability
condition, upon making use of (\ref{f1norm}), leads to
\begin{equation}
  \begin{array}{ll}
    g_2 & =2\left(\psi_{1\xi}(\xi_0+\epsilon\xi_1)+
      \psi_{0\xi}(\xi_0+\epsilon\xi_1)-
      \psi_{0\xi}(\xi_0)\right) \\ & =2\psi_{1\xi}(\xi_0)+\Ord{\epsilon} \\
    & \approx0.06354624322+\Ord{\epsilon} \nonumber,
  \end{array}
  \label{g2}
\end{equation}
where the even symmetry of $\psi_0$ and $\psi_1$ is taken into
account.  The numerical results given in (\ref{xi1}) and (\ref{g2})
coincide with the ones reported by Cole and Wagner in their paper
\cite[p.167]{ColeWag} though our values are given with more
significant digits.  Consequently, the next order centre manifold
model is
\begin{equation}
  \dot a \approx a(\epsilon g_1+\epsilon^2 g_2)
  \label{eveq}
\end{equation}
with solution
\begin{equation}
        a=a_0t^{-\alpha'/2}\,,
        \quad\mbox{where}\quad
        \alpha'=2\epsilon\left(\frac{1}{\sqrt{2\pi e}}-
        \epsilon g_2\right).
        \label{A2}
\end{equation}

\section[The correct initial condition ensures fidelity]%
{The correct initial condition ensures fidelity of the model}

The correct projection of initial conditions onto a centre manifold,
first developed in \cite{Roberts89b} and recently refined in
\cite{Roberts97b}, should approximately determine the ``functional of
the initial conditions'' mentioned by Barenblatt near the top of p.202
\cite{Barenblatt96}, but not previously found.
Here we follow the procedure outlined in \cite{Roberts97b} to give the
proper initial conditions $a_0$ for the centre manifold model
(\ref{A2}) when the initial conditions for the original problem are
given by $\theta=\theta_0(x)$ at $t=1$ corresponding to $\tau=0$.
We expect that $a|_{\tau=0}=\int_{-\infty}^{\infty}\theta_0\,dx$, but
this is only a first approximation.
The more careful analysis corrects this approximation.

As used in previous sections, the special form of (\ref{Ephit}) implies that
its solution is to be found in the separable form
\begin{equation}
  \phi(\tau,\xi;\epsilon)=a(\tau)\psi(\xi;\epsilon),\quad\mbox{where}\quad
  \dot{a}=a(\tau)g(\epsilon).
\end{equation}
Then ``vectors'' locally tangent to the centre manifold are found to be
${\bf e}_1=(a\partial\psi/\partial\epsilon,1)$ and ${\bf e}_2=(\psi,0)$.
According to \cite{Roberts97b} we need to find ``vectors'' ${\bf z}_1$ and
${\bf z}_2$ satisfying
\begin{equation}
  {\cal D} {\bf z}_j-
  \sum_{k=1}^2\left\langle {\cal D}{\bf z}_j,{\bf e}_k \right\rangle{\bf
z}_k={\bf0}\,,\quad j=1,2
  \label{Dz}
\end{equation}
and normalisation $\left\langle{\bf z}_j,{\bf e}_k \right\rangle=\delta_{jk}$
where the dual operator $\cal D$ is defined as
\begin{equation}
  {\cal D}\equiv \frac{\partial}{\partial\tau}+{\cal I}^\dag\,,
\end{equation}
the adjoint
\begin{equation}
  {\cal I}^\dag=\left[ \begin{array}{cc}
      {\cal L}^\dag+\epsilon D_{\bar{\xi}}^{\dag}&0\\
      \left.D_{\bar{\xi}}\phi
      +\epsilon D_{\bar{\xi}}
      \frac{\partial\phi}{\partial\epsilon}\right.
      & 0
    \end{array} \right]
  \label{adji}
\end{equation}
and
\begin{equation}
      D_{\bar{\xi}}^{\dag}\equiv
      D_{\bar{\xi}}
      +2\left(\delta(\xi+\bar{\xi})-\delta(\xi-\bar{\xi})\right)
      \frac{\partial}{\partial\xi}
      +\delta^\prime(\xi+\bar{\xi})-\delta^\prime(\xi-\bar{\xi})\,,
\end{equation}
in which $\delta$ and $\delta^\prime$ denote the Dirac delta function and its
derivative, respectively.
The normalisation conditions give that
\begin{equation}
  \left[ \begin{array}{c}
      z_1^{(1)}\\z_1^{(2)}
    \end{array} \right] =
  \left[ \begin{array}{c}
      \frac{1}{a}r_1^{(1)}(\xi)\\r_1^{(2)}(\xi)
    \end{array} \right] ,\quad
  \left[ \begin{array}{c}
      z_2^{(1)}\\z_2^{(2)}
    \end{array} \right] =
  \left[ \begin{array}{c}
      r_2^{(1)}(\xi)\\ar_2^{(2)}(\xi)
    \end{array} \right] ,
\end{equation}
\begin{equation}
  \begin{array}{l}
    \int_{-\infty}^\infty r_1^{(1)}\psi\,d\xi=0,\quad
    \int_{-\infty}^\infty \left(r_1^{(1)}\frac{\partial\psi}{\partial\epsilon}
    +r_1^{(2)}\right)\,d\xi=1,\\
    \int_{-\infty}^\infty r_2^{(1)}\psi\,d\xi=1,\quad
    \int_{-\infty}^\infty \left(r_2^{(1)}\frac{\partial\psi}{\partial\epsilon}
    +r_2^{(2)}\right)\,d\xi=0
  \end{array}.
  \label{nrm}
\end{equation}

We look for the solution of (\ref{Dz}) satisfying ${\cal D}{\bf z}_1={\bf0}$,
{\em i.e.}
\begin{equation}
  \left[ \begin{array}{c}
      -\frac{g}{a}r_1^{(1)}\\0
    \end{array} \right]=-
  \left[ \begin{array}{c}
      \frac{1}{a}\left({\cal L}^\dag
      +\epsilon D_{\bar{\xi}}\right)r_1^{(1)}\\
      \left(D_{\bar{\xi}}\psi
      +\epsilon D_{\bar{\xi}}
      \frac{\partial\psi}{\partial\epsilon}\right)r_1^{(1)}
    \end{array} \right].
\end{equation}
Hence we immediately deduce that $r_1^{(1)}=0$. Consequently, the second of
normalisation conditions (\ref{nrm}) is transformed to
$\int_{-\infty}^\infty r_1^{(2)} d\xi=1$. Then from the projection of initial
conditions
\begin{equation}
  \frac{1}{a|_{\tau=0}}\left\langle r_1^{(1)},\theta_0-a|_{\tau=0}\psi
\right\rangle
  +(\epsilon_0-\epsilon) \left\langle r_1^{(2)},1 \right\rangle=0
\end{equation}
and we deduce that $\epsilon\equiv\epsilon_0$.
This result, that the parameter $\epsilon$ remains unchanged between
the model and the original problem, is expected at the outset, but we
have just demonstrated how it is obtained in the context of the
developed theory for the projection of initial conditions.

Thus the proper initial condition for the amplitude $a|_{\tau=0}$ is given by
\begin{equation}
  \left\langle r_2^{(1)},\theta_0-a|_{\tau=0}\psi \right\rangle=0,
\end{equation}
 or, equivalently, since the problem is linear in amplitude $a$ and the
 normalisation conditions (\ref{nrm}) are used, by
\begin{equation}
  a|_{\tau=0}=\left\langle r_2^{(1)},\theta_0 \right\rangle.
  \label{u0}
\end{equation}
Thus the problem of finding the proper initial condition is reduced to solving
for $r_2^{(1)}$ which satisfies the following equation deduced from
(\ref{Dz})
\begin{equation}
  \left({\cal L}^\dag+\epsilon  D_{\bar{\xi}}^{\dag}
  \right)r_2^{(1)}
  =\left \langle
    \left({\cal L}^\dag+\epsilon D_{\bar{\xi}}^{\dag}\right)
    r_2^{(1)},\psi
  \right\rangle r_2^{(1)}\,.
  \label{41}
\end{equation}
Performing integration by parts in the right-hand side of (\ref{41}) and
using the normalisation (\ref{nrm}) we obtain
\begin{equation}
  \left({\cal L}^\dag+\epsilon D_{\bar{\xi}}^{\dag}\right)
  r_2^{(1)}-gr_2^{(1)}=0,\quad \left \langle r_2^{(1)},\psi\right\rangle =1.
  \label{r2}
\end{equation}
We solve (\ref{r2}) assuming
$r_2^{(1)}=p_0(\xi)+\epsilon p_1(\xi)+\Ord{\epsilon^2}$ and recollecting that
$g\approx-\epsilon/\sqrt{2 \pi e}+\Ord{\epsilon^2}$ and
$\psi=\psi_0+\epsilon\psi_1+\Ord{\epsilon^2}$. At $\Ord{\epsilon^0}$ we obtain
\begin{equation}
    {\cal L}^\dag p_0=0, \quad \left \langle p_0,\psi_0\right\rangle =1
\end{equation}
with solution $p_0=z=1$. Thus at leading order $a|_{\tau=0}=\int_{-\infty}
^\infty \theta_0(\xi)d\xi$.

\begin{figure}[tbp]
  \centerline{\includegraphics{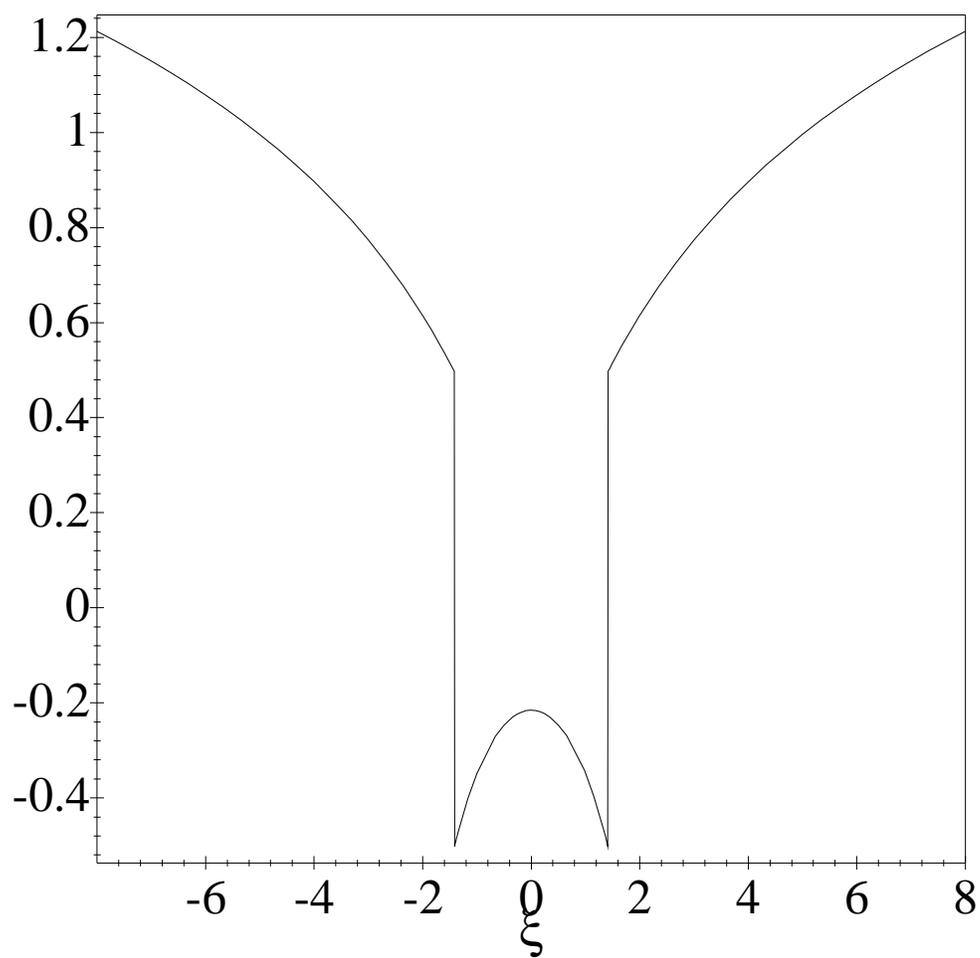}}
  \caption{$\Ord{\epsilon}$ initial condition projection function $p_1(\xi)$.}
  \label{f2}
\end{figure}

At $\Ord{\epsilon^1}$ we obtain
\begin{equation}
    {\cal L}^\dag p_1+\delta^\prime(\xi+\xi_0)-\delta^\prime(\xi-\xi_0)
    +\frac{1}{\sqrt{2 \pi e}}=0,
    \quad \left \langle p_1,\psi_0\right\rangle=0.
\end{equation}
The solution, presented in Figure~\ref{f2}, has the following
algebraic form
\begin{equation}
  \begin{array}{rl}

p_1(\xi)=&c_4+\left(1+i\sqrt{\frac{\pi}{2e}}\,\mbox{erf}\left(\frac{i}{\sqrt
2}\right) \right)
    (H(\xi-\xi_0)-H(\xi+\xi_0))\\&
    -\frac{i}{\sqrt{2e}}\int_0^{\xi}
    \mbox{erf}\left(\frac{iy}{2}\right)e^{-y^2/4}dy\\&
    +i\sqrt{\frac{\pi}{2e}}\left(1+
      \mbox{erf}(\frac{\xi}{2})-H(\xi+\xi_0)-H(\xi-\xi_0)\right)\,
    \mbox{erf}\left(\frac{i\xi}{2}\right),
  \end{array}
  \label{p1}
\end{equation}
where
\begin{equation}
  \begin{array}{rl}
   c_4=&\frac{i}{\sqrt{2\pi e}}(I_3-\sqrt{\pi}(I_2+I_1))
       +\left(1+i\sqrt{\frac{\pi}{2e}}\,
         \mbox{erf}\left(\frac{i}{\sqrt2}\right)\right)\,
       \mbox{erf}\left(\frac{1}{\sqrt2}\right)\\
       \approx & 0.0589390531\,.
 \end{array}
\end{equation}

Finally we then have that the proper initial condition for the centre manifold
model~(\ref{eveq}) is given by
\begin{equation}
  a|_{\tau=0}=\int_{-\infty}^{\infty}\left(1+\epsilon p_1(\xi)
    +\Ord{\epsilon^2}\right)\theta_0
  \left(\xi\right)\,d\xi
  \label{ic}
\end{equation}
Note that $p_1\sim[2/(\pi e)]^{1/2}\log(|\xi|)$ as $|\xi|\to\infty$
and, consequently, the integral~(\ref{u0}) converges only for
a sufficiently compact initial distribution $\theta_0$.
This emphasises that the projection of the initial conditions is local
in its nature and it is applicable only if the initial conditions for
the original problem are, in some sense, close to the centre manifold.

\section{Choose an optimal origin in time and space}

It follows from the transformation of space and time variables
(\ref{Etrans}) that the diffusion from a localised initial release of
arbitrary form occurring in the original problem at $t=1$ is modelled
by the evolution from the initial state of a point release, a delta
function, at $x=t=0$.
On the other hand the original partial differential equation
(\ref{Ebaren}) is invariant with respect to translations in time and
space.
Thus there is freedom to choose the time and space origins for the
model to suit best the actual distribution of the initial $\theta$.
To account for these inherit degrees of freedom in the original
problem we generalise the coordinate transformation (\ref{Etrans}) to
\begin{equation}
        \tau=\log {(t+t_0)}\,,\quad
        \xi=\frac{x-x_0}{\sqrt{t+t_0}}\,,\quad
        \theta=\frac{\phi(\tau,\xi)}{\sqrt{t+t_0}}\,,
        \label{Egentran}
\end{equation}
where $t_0>0$.
Now the localised release $\theta_0(x)$ occurring in the original
problem at time $t=0$ (not at $t=1$ as assumed in the previous
sections) is modelled by some Gaussian centred at $x_0$ rather than by
the delta function at $x=0$.
The width of the model Gaussian at the moment of the actual release
$t=0$ is determined by $t_0$ which also determines the location of the
virtual origin in time for the model.
Generalisation~(\ref{Egentran}) does not affect the analysis of the
previous sections.
In particular, the model dynamics (\ref{eveq}) is unchanged because
the general long-term dynamics are independent of the space-time
origin.
However, the generalisation provides a two-parameter family of model
solutions to the original problem (\ref{Ebaren}) rather than just the
unique model described earlier.
Thus here the general projection of initial condition (\ref{ic}) becomes
\begin{equation}
  a_0=t_0^{\alpha^\prime/2}\int_{-\infty}^{\infty}
  \left[1+\epsilon p_1\left(\frac{x-x_0}{\sqrt{t_0}}\right)
    +\Ord{\epsilon^2}\right]\theta_0(x)\,dx\,.
  \label{ic1}
\end{equation}
One is free to choose parameters $x_0$ and $t_0$ entering (\ref{ic1})
in such a way that the model possess certain additional properties.
For instance, we choose $t_0$ such that the contribution of the
$\epsilon$-dependent terms in (\ref{ic1}) is zero---this choice should
ensure that the model $a$ most closely matches the solution $\theta$
for the original problem in the short-term as well as the long-term
evolution.
In essence this is equivalent to considering all the centre manifolds
(in $a$ and $\epsilon$) parameterized by $t_0$ and $x_0$, and choosing
that centre manifold whose isochrons are linearly ``vertical'' and
hence make the definition of $a$ match the projection.
It is always possible to make this choice since physical initial
distributions $\theta_0$ are non-negative functions while the mean of
$p_1$ is zero.
Thus require
\begin{equation}
  I=\int_{-\infty}^{\infty}p_1\left(\frac{x-x_0}{\sqrt{t_0}}\right)
  \theta_0(x)\,dx=0\,
  \label{t0}
\end{equation}
which we view as implicitly defining $t_0$ as a function of $x_0$.

The value of $x_0$ is then fixed to minimise $t_0$.
We feel this is desirable since it minimises the spread of the model's
Gaussian at the initial instant of release and so maximises the
information content of the model.
(It is also the only distinguished $x_0$.) Differentiating (\ref{t0})
with respect to $x_0$ we obtain
\begin{equation}
  \frac{dI}{dx_0}=-\frac{1}{\sqrt t_0}\int_{-\infty}^{\infty}p_1^\prime
  \left(\frac{x-x_0}{\sqrt{t_0}}\right)\theta_0(x)
  \left[1+\frac{x-x_0}{2t_0}\frac{dt_0}{dx_0}\right]\,dx=0\,,
  \label{Iprime}
\end{equation}
where prime denotes differentiation with respect to the argument.
At the point of extremum $dt_0/dx_0=0$ and the second term in the
brackets in (\ref{Iprime}) vanishes.
Thus we solve
\begin{equation}
  \int_{-\infty}^{\infty}p_1^\prime
  \left(\frac{x-x_0}{\sqrt{t_0}}\right)\theta_0(x)\,dx=0\,.
  \label{x0}
\end{equation}
in conjunction with (\ref{t0}) to define $x_0$ and $t_0$.
As an aside it follows from the above discussion that such chosen
$x_0$ and $t_0$ guarantee that $I=0$ is a minimum contribution to the
$\epsilon$-correction of initial conditions for the model.
If $\theta_0$ is symmetric, say about $x=q$, then, owing to the even
symmetry of $p_1$, the choice of $x_0=q$ guarantees that (\ref{x0}) is
satisfied.  Thus for symmetric $\theta_0$ the best choice for the
centre of the Gaussian spread of the model is the point of symmetry.

Finally, the initial amplitude is then given by
\begin{equation}
  a_0=t_0^{\alpha'/2}\int_{-\infty}^\infty\theta_0(x)\,dx
\end{equation}
and the model solution written in the original variables becomes
\begin{equation}
  \theta=\frac{a_0}{(t+t_0)^{(1+\alpha^\prime)/2}}
  \left[\psi_0\left(\frac{x-x_0}{\sqrt{t+t_0}}\right)
    +\epsilon\psi_1\left(\frac{x-x_0}{\sqrt{t+t_0}}\right)
    +\Ord{\epsilon^2}\right]\,,
  \label{theta}
\end{equation}
where $t_0$ and $x_0$ satisfy (\ref{t0}) and (\ref{x0}).

\section[Numerical results demonstrate the accuracy]%
{Numerical results demonstrate the accuracy of the model}

We illustrate the correctness of the derived initial conditions by
comparing the model predictions with the direct numerical integration
of equation (\ref{Ebaren}).
Let the initial distribution of substance for the original problem at
$t=0$ be in the form of the Gaussian
\begin{equation}
  \theta_0=\sqrt{\frac{10}{\pi}}\exp(-10x^2)\,.
  \label{nic}
\end{equation}

Numerical integration of (\ref{Ebaren}) with initial distribution
(\ref{nic}) was performed using \textsc{imsl} routine \textsc{dmolch}
\cite{IMSL} with the accuracy of $10^{-8}$.
Since the long term behaviour of the numerical solution was found to
depend on the size of the computational domain, the preliminary test
of the numerical solution was performed for $\epsilon=0$ for which the
analytic solution comes from (\ref{Egauss}).
It was found that the non-reflecting boundary conditions
$\theta_x(L)/\theta(L)=x/(2t)$ imposed at $L=22.5$ eliminated such an
influence for the time interval considered.

\begin{figure}[tbp]
  \centerline{\includegraphics{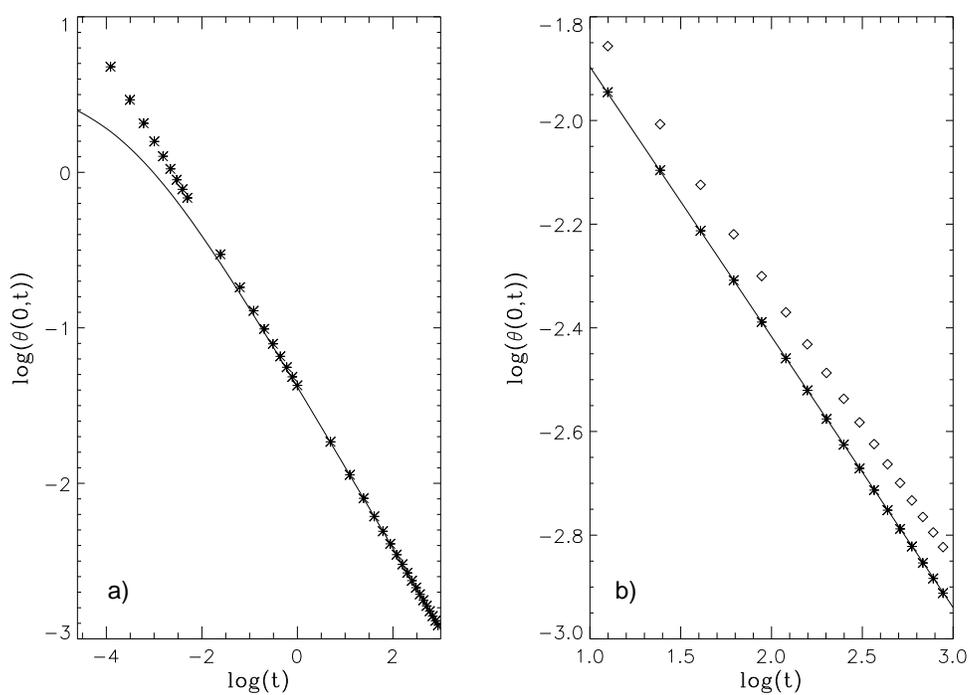}}
  \caption{Numerical (solid line) solutions of
    equation~(\ref{Ebaren}) evaluated at $x=0$ for $\epsilon=0.1$
    compared with the model~(\ref{theta}) that uses the correct
    initial conditions (stars) and the previous model~(\ref{eam})
    (diamonds).}
  \label{f3}
\end{figure}

The resulting time evolution of the direct numerical solution for
$\epsilon=0.1$ at $x=0$ is shown by a solid line in Figure~\ref{f3}.
Because of the symmetry of initial distribution (\ref{nic}) with
respect to the line $x=0$, (\ref{x0}) gives the value $x_0=0$ for
model (\ref{theta}).
Numerical evaluation shows that condition (\ref{t0}) is satisfied for
$\theta_0$ given by (\ref{nic}) for $t_0\approx0.0250$.
As seen from Figure~\ref{f3}(a) the model dynamics shown by star
symbols approaches the numerical curve very quickly.
In Figure~\ref{f3}(b) we compare the numerical and the proper
model~(\ref{theta}) solutions with the earlier proposed model
\cite{Barenblatt96,ColeWag}
\begin{equation}
  \theta=\frac{\int_{-\infty}^\infty\theta_0(x)\,dx}
  {t^{(1+\alpha^\prime)/2}}
  \left(\psi_0\left(\frac{x}{\sqrt t}\right)+
    \epsilon\psi_1\left(\frac{x}{\sqrt t}\right)+\Ord{\epsilon^2}\right)
  \label{eam}
\end{equation}
which uses naive initial condition (\ref{Enaiveic})---shown by diamond
symbols.
While the present model and numerical solution are virtually
indistinguishable in their long term evolution, model (\ref{eam})
based solely on scaling arguments is able to predict just a slope.
The actual values of the distribution maximum it provides lies apart
from the numerical curve for all time.
Thus the correct initial conditions for the model are essential to
avoid a permanent finite phase difference between the model and
the actual full solutions.

\section{Conclusions}

We have demonstrated that the centre manifold theory provides a
straightforward and rigorous way of deriving not only the functional
form of similarity solutions of nonlinear diffusion, but also the
appropriate initial conditions for the model in terms of the initial
distributions of the substance.
This cannot be done using other modelling approaches such as, for
example, scaling laws or the method of multiple scales.
The correct provision of initial conditions also enables us to
determine an optimal location for the virtual space-time origin for
the model.
The present technique may be successfully used for modelling a wide
class of nonlinear filtration/diffusion problems.

\end{document}